\documentclass[aps,showpacs,prb,reprint,superscriptaddress]{revtex4-1}

\usepackage{hyperref}
\usepackage{setspace} 
\usepackage{graphicx}
\usepackage{amsmath}
\usepackage{color}
\usepackage{amsmath}
\usepackage{amssymb}
\usepackage{verbatim}
\usepackage{latexsym}
\usepackage{enumerate} 
\usepackage{bm} 

\begin{document}

\title{Anharmonic vibrational properties in periodic systems: energy,
  electron-phonon coupling, and stress}

\author{Bartomeu Monserrat}

\affiliation{TCM Group, Cavendish Laboratory, University of Cambridge,
  J.\ J.\ Thomson Avenue, Cambridge CB3 0HE, United Kingdom}

\author{N.\ D.\ Drummond}

\affiliation{Department of Physics, Lancaster University, Lancaster
  LA1 4YB, United Kingdom}

\author{R.\ J.\ Needs}

\affiliation{TCM Group, Cavendish Laboratory, University of Cambridge,
  J.\ J.\ Thomson Avenue, Cambridge CB3 0HE, United Kingdom}

\date{\today}

\begin{abstract}
  A unified approach is used to study vibrational properties of periodic
  systems with first-principles methods and including anharmonic effects.
  Our approach provides a theoretical basis for the determination of
  phonon-dependent quantities at finite temperatures.  The low-energy portion
  of the Born-Oppenheimer energy surface is mapped and used to calculate the
  total vibrational energy including anharmonic effects, electron-phonon
  coupling, and the vibrational contribution to the stress tensor.  We report
  results for the temperature dependence of the electronic band gap and the
  linear coefficient of thermal expansion of diamond, lithium hydride, and
  lithium deuteride.
\end{abstract}


\maketitle



\section{Introduction}

The atomic vibrations that are always present in condensed matter lead to important corrections to the static-lattice model of crystalline solids. The most common approach for evaluating such corrections is to include vibrational effects within the harmonic phonon approximation,\cite*{wallace,born,maradudin} which works well in many circumstances. However, anharmonic effects can be important in, for example, systems incorporating light atoms which have large vibrational amplitudes, systems with weak bonding such as hydrogen bonding, and systems at high temperatures. When the vibrations are present, their effect on the electrons can also become important. Electron-phonon interactions are central, for example, to the description of superconductivity.

Renormalization of the phonon degrees of
freedom is an important idea underlying most treatments of phonon-phonon and
electron-phonon interactions. For example: (i) the phonon frequencies as a
function of temperature may be used to investigate the stability of crystals
with unstable harmonic phonons, (ii) the change in phonon frequencies due to
electron-phonon interactions may be used in the study of the
temperature dependence of electronic band gaps, and (iii) the phonon
frequencies at different volumes may be used within the quasiharmonic
approximation to study thermal expansion.

The self-consistent phonon method was devised by Born and
Hooton,\cite{born_self,hooton,born_hooton,doi:10.1080/14786435808243224} and
extended by Choquard\cite{choquard} and Werthamer.\cite{PhysRevB.1.572} In the self-consistent phonon method, the phonon frequencies are renormalized by considering the atomic vibrations over a region about equilibrium which can include anharmonicity. For example, it is useful in systems with a harmonic instability, such as solid helium which forms under pressure, where including anharmonic terms leads to the correct physical picture in which the body-centered-cubic structure is stabilized. A historical overview of these developments is given by Klein and Horton.\cite{klein}


More recently, these ideas have been used in first-principles density
functional theory (DFT) electronic structure calculations and have shown the
occurrence of dynamically unstable phases at the harmonic level that are
stabilized at finite temperature by means of
anharmonicity.\cite{RevModPhys.84.945}
Souvatzis and co-workers,\cite{PhysRevLett.100.095901} and Hellman and
co-workers\cite{PhysRevB.84.180301} have calculated renormalized phonons at
finite temperature for systems with harmonic phonon instabilities. Antolin and
co-workers\cite{PhysRevB.86.054119} presented a simpler and computationally
less demanding method for solving this problem, using a harmonic fit to
large-amplitude phonon displacements.

Thermal expansion arises from anharmonicity and is usually addressed within
the quasiharmonic approximation,\cite{PhysRevB.71.205214} in which the phonon
frequencies are taken to depend on the volume. The minimization of the free
energy with respect to volume at a given temperature leads to the equilibrium
volume at that temperature. This method has been very successful, but for
systems with several lattice parameters it is computationally expensive to
relax the structures at each volume.

Lattice vibrations also play a central role in electron-phonon
interactions. The study of electron-phonon interactions within many-body
perturbation theory is well established.\cite{mahan} The effects of
electron-phonon interactions on the temperature dependence of band gaps are
substantial, and they have been described within the harmonic
approximation.\cite*{RevModPhys.77.1173,0022-3719-9-12-013,PhysRevB.23.1495} Recent theoretical advances have enabled such calculations to be performed from first principles.\cite{PhysRevLett.105.265501}

In this paper we describe a unified approach for studying the effects of
anharmonicity at zero and finite temperatures on quantities including phase
stability, the total energy, electron-phonon coupling leading to band gap
renormalization, and thermal expansion.


The starting point of our method involves performing first-principles
calculations of the low-energy portion of the Born-Oppenheimer
(BO)\cite{bornoppenheimer} energy surface defined by the harmonic phonon
coordinates, and exploring it out to large atomic displacements where the
harmonic approximation is no longer accurate.  This information is used to
calculate anharmonic phonon free energies and the associated vibrational
excited states using the vibrational self-consistent-field (VSCF)
equations.\cite{bowman:608} We then use a perturbation theory constructed on
the VSCF equations\cite{norris:11261} to obtain a second-order correction to
the total vibrational free energy. This method has been used before in
studying molecules\cite*{bowman:608,jung:10332} and extended model
systems,\cite{PhysRevB.78.224108} but, as far as we are aware, this is the
first first-principles application to three-dimensional periodic solids.

We have used the anharmonic wave functions to calculate quantum-mechanical
expectation values of phonon-dependent quantities at zero and finite
temperatures. For example, renormalization of the electronic eigenvalues due
to their interaction with the phonons leads to the temperature dependence of
the electronic band structure. Similarly, the vibrational stress tensor can be
calculated and used to directly study thermal expansion.

The main computational cost in our approach is that of performing the electronic structure calculations to explore the BO energy surface. While performing these calculations, data for all quantities of interest can be accumulated, and the further analysis of particular quantities can be done with little additional computational cost.

We report results for the anharmonic energy, the temperature dependence of the
thermal band gap, and the linear coefficient of thermal expansion of diamond,
lithium hydride (H$^7$Li) and lithium deuteride (D$^7$Li). Our results are in
good agreement with experiment and other theoretical approaches. As far as we
are aware, no experimental or theoretical results for the temperature
dependence of the band gap of lithium hydride have been published previously.

The remainder of this paper is arranged as follows. In
Sec.\ \ref{sec:anh_theory} we present the theoretical framework used to study
anharmonicity and in Sec.\ \ref{sec:ph_theory} we extend the formalism to the
evaluation of general phonon expectation values. In Sec.\ \ref{sec:comp} we
describe our computational approach. Our results are presented in
Sec.\ \ref{sec:results} and we draw our conclusions in
Sec.\ \ref{sec:conclusions}. All equations are given in Hartree atomic units,
in which the Dirac constant, the electronic charge and mass, and $4\pi$ times
the permittivity of free space are unity ($\hbar=|e|=m_{\mathrm{e}}=4\pi
\epsilon_0=1$).

\section{Anharmonic total energy} \label{sec:anh_theory}

\subsection{BO approximation}

Within the BO approximation\cite{bornoppenheimer} the electronic and
vibrational degrees of freedom are treated separately.
The atomic positions parameterize a family of electronic Hamiltonians
$\hat{H}_{\mathrm{el}}(\mathbf{R})$ labeled by the collective nuclear position
variable $\mathbf{R}$. Each of these Hamiltonians leads to the electronic
energy $E_{\mathrm{el}}$ and wave function $|\Psi\rangle$ according to
$\hat{H}_{\mathrm{el}}(\mathbf{R})|\Psi(\mathbf{R})\rangle=E_{\mathrm{el}}(\mathbf{R})|\Psi(\mathbf{R})\rangle$.
In the following we consider only the electronic ground state.

The nuclear or vibrational Hamiltonian $\hat{H}_{\mathrm{vib}}$ is
\begin{equation}
\hat{H}_{\mathrm{vib}}=\sum_{\mathbf{R}_p,\alpha}-\frac{1}{2m_{\alpha}}\nabla^2_{p\alpha}+E_{\mathrm{el}}(\mathbf{R}),
\end{equation}
where the $\mathbf{R}_p$ are the position vectors of the unit cells that make
up the periodic supercell, $\alpha$ labels the different atoms within a unit
cell, and $m_{\alpha}$ is the mass of atom $\alpha$. In this context, the
electronic eigenvalue as a function of atomic position
$E_{\mathrm{el}}(\mathbf{R})$ is called the BO surface.

\subsection{Harmonic approximation}

Within the harmonic approximation (HA)\cite*{wallace,born,maradudin} the BO
energy surface is approximated as a quadratic function of the atomic
displacement coordinates
$\mathbf{u}_{p\alpha}=\mathbf{r}_{p\alpha}-\mathbf{r}^0_{p\alpha}$,
\begin{eqnarray}
  \hat{H}_{\mathrm{vib}}&=&\sum_{\mathbf{R}_p,\alpha}-\frac{1}{2m_{\alpha}}\nabla_{p\alpha}^2+E_{\mathrm{el}}(\mathbf{R}^0)  \nonumber \\
  &+&\,\frac{1}{2}\sum_{\substack{\mathbf{R}_p,\alpha,i\\\mathbf{R}_{p'},\alpha',j}}\frac{\partial^2E_{\mathrm{el}}(\mathbf{R}^0)}{\partial\,u_{p\alpha;i}\,\partial\,u_{p'\alpha';j}}u_{p\alpha;i}\,u_{p'\alpha';j}\,, \label{eq:disp_hamil}
\end{eqnarray}
where $\mathbf{r}_{p\alpha}$ are the atomic positions,
$\mathbf{r}^0_{p\alpha}$ the equilibrium atomic positions, and Latin indices
$i$ and $j$ are used to label Cartesian directions.  The HA often works very
well because the nuclei are heavy and therefore only explore the neighborhood
of their equilibrium positions. In the following we drop the constant energy
term $E_{\mathrm{el}}(\mathbf{R}^0)$.

The matrix of force constants is defined as $D_{i\alpha;j\alpha'}(\mathbf{R}_p,\mathbf{R}_p')=\partial^2E_{\mathrm{el}}(\mathbf{R}^0)/\partial\,u_{p\alpha;i}\,\partial\,u_{p'\alpha';j}$, and the dynamical matrix at the reciprocal space point $\mathbf{k}$ is
\begin{equation}
D_{i\alpha;j\alpha'}(\mathbf{k})\!=\!\frac{1}{N_p\sqrt{m_{\alpha}m_{\alpha'}}}\!\!\sum_{\mathbf{R}_p,\mathbf{R}_{p'}}\!\!\!\!D_{i\alpha;j\alpha'}(\mathbf{R}_p,\mathbf{R}_p')e^{i\mathbf{k}\cdot(\mathbf{R}_p-\mathbf{R}_p')} ,
\end{equation}
where $N_p$ is the number of unit cells within the periodic supercell.

The vibrational Hamiltonian of Eq.\ (\ref{eq:disp_hamil}) can be rewritten in
terms of normal or phonon coordinates $q_{n\mathbf{k}}$, which are related to
displacement coordinates by
\begin{eqnarray}
u_{p\alpha;i}&=&\frac{1}{\sqrt{N_pm_{\alpha}}}\sum_{n,\mathbf{k}}q_{n\mathbf{k}}e^{i\mathbf{k}\cdot\mathbf{R}_p}w_{\mathbf{k}n;i\alpha}, \label{eq:at_amp} \\
q_{n\mathbf{k}}&=&\frac{1}{\sqrt{N_p}}\sum_{\mathbf{R}_p,\alpha,i}\sqrt{m_{\alpha}}\,u_{p\alpha;i}e^{-i\mathbf{k}\cdot\mathbf{R}_p}w_{-\mathbf{k}n;i\alpha}. \label{eq:ph_amp}
\end{eqnarray}
Here, $w_{\mathbf{k}n;i\alpha}$ are the eigenvectors of the dynamical matrix and $n$ is the phonon branch index. In terms of phonon coordinates, which can always be chosen to be real,\cite{maradudin} the vibrational Hamiltonian is
\begin{equation}
\hat{H}_{\mathrm{vib}}=\sum_{n,\mathbf{k}}-\frac{1}{2}\frac{\partial^2}{\partial q_{n\mathbf{k}}^2}+\frac{1}{2}\omega_{n\mathbf{k}}^2q_{n\mathbf{k}}^2, \label{eq:harm_hamil}
\end{equation}
where $\omega^2_{n\mathbf{k}}$ are the eigenvalues of the dynamical
matrix. This Hamiltonian consists of a sum of terms for non-interacting simple
harmonic oscillators of frequencies $\omega_{n\mathbf{k}}$.

\subsection{Anharmonic approximation}

\subsubsection{Principal axes approximation to the BO energy
  surface}

To study anharmonic properties it is necessary to find an approximation to the
BO energy surface that goes beyond the HA\@. The HA is often very accurate,
and in such cases anharmonicity is expected to be a perturbation about the
harmonic case. We write the BO energy surface as\cite{jung:10332}

\begin{eqnarray}
E_{\mathrm{el}}(\mathbf{Q})&=&E_{\mathrm{el}}(\mathbf{0})+\sum_{n,\mathbf{k}}V_{n\mathbf{k}}(q_{n\mathbf{k}}) \nonumber \\
&+&\frac{1}{2}\sum_{n,\mathbf{k}}\sum_{n',\mathbf{k}'}\!{}^{'}V_{n\mathbf{k};n'\mathbf{k}'}(q_{n\mathbf{k}},q_{n'\mathbf{k}'})+\cdots,
\end{eqnarray}
where $\mathbf{Q}$ is a collective phonon vector with elements $q_{n\mathbf{k}}$, the primed sum indicates that the term $(n,\mathbf{k})=(n',\mathbf{k}')$ is excluded and the factor $1/2$ accounts for double counting. The independent phonon term takes the form
\begin{equation}
V_{n\mathbf{k}}(q_{n\mathbf{k}})=E_{\mathrm{el}}(0,\ldots,q_{n\mathbf{k}},\ldots,0)-E_{\mathrm{el}}(\mathbf{0}), \label{eq:indep_term}
\end{equation}
but note that we do not assume the HA\@.  The two-body term which introduces
coupling between the modes takes the form 
\begin{eqnarray}
V_{n\mathbf{k};n'\mathbf{k}'}&&(q_{n\mathbf{k}},q_{n'\mathbf{k}'})\!=\!E_{\mathrm{el}}(0,\ldots,q_{n\mathbf{k}},\ldots,q_{n'\mathbf{k}'},\ldots,0) \nonumber \\
&&-V_{n\mathbf{k}}(q_{n\mathbf{k}})\!-\!V_{n'\mathbf{k}'}(q_{n'\mathbf{k}'})\!-\!E_{\mathrm{el}}(\mathbf{0}). \label{eq:coup_term}
\end{eqnarray}
It is possible to continue this series for more general $N$-body terms, but
because we start from the HA in which phonons are exactly noninteracting, it
is expected that the higher-order terms will decrease in magnitude as $N$
increases.
[We note that Eq.\ (14) of Ref.\ \onlinecite{jung:10332} contains a
typographical error connected with the coupling of vibrational modes.]

\begin{figure}
\includegraphics[scale=0.35]{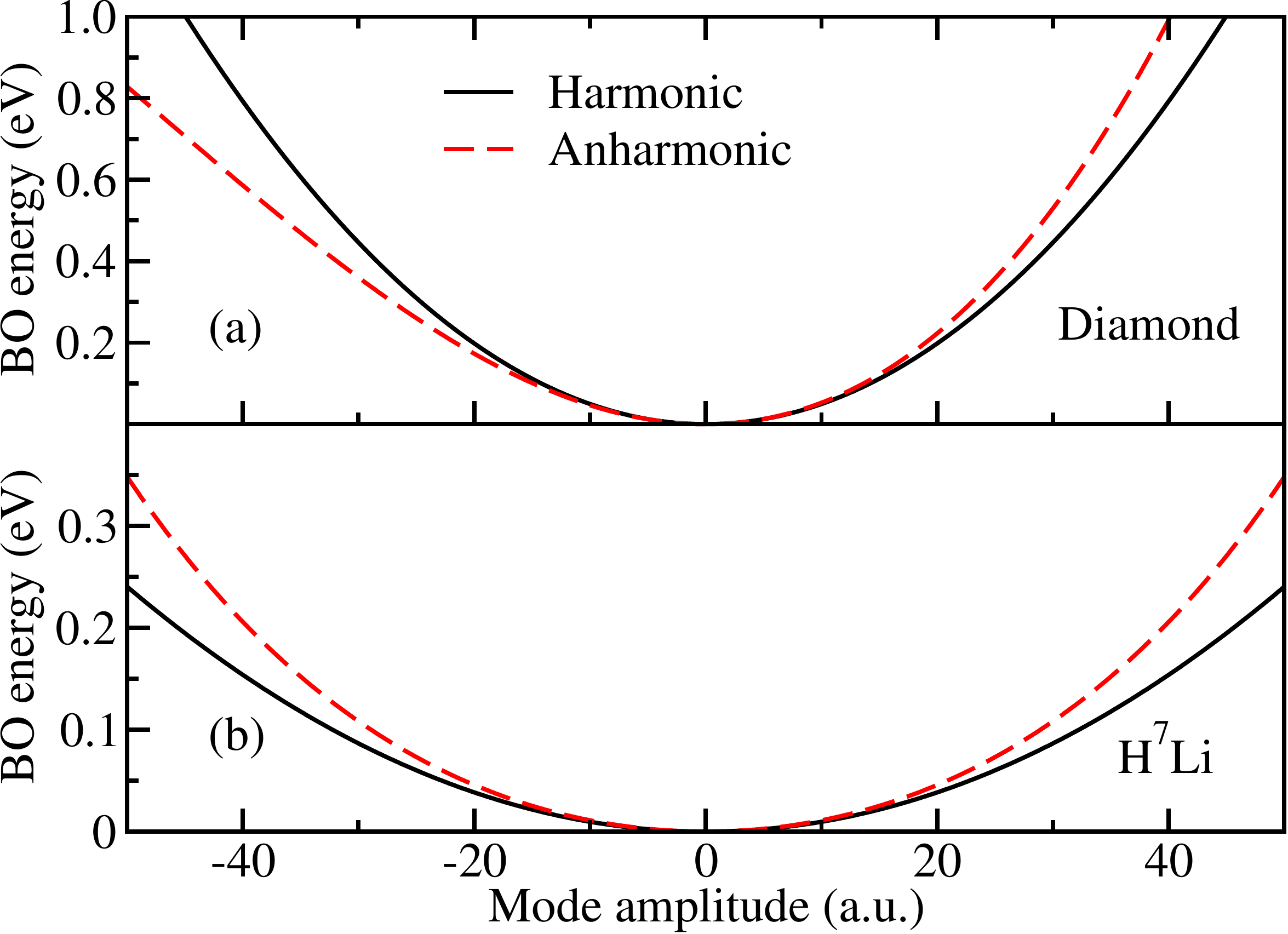}
\caption{(Color online) Harmonic and anharmonic BO energy surfaces per unit
  cell for an optical $\Gamma$-point phonon in (a) diamond and (b)
  H$^7$Li. The leading anharmonic term for diamond is cubic, and that for
  H$^7$Li is quartic. For a harmonic ground-state wave function (Gaussian),
  the one-sigma amplitude is $9.1$ a.u.\ for diamond and $13.7$ a.u.\ for
  H$^7$Li. In real space, the one-sigma amplitudes correspond to 0.043
  a.u.\ for diamond and 0.299 a.u.\ for H$^7$Li.}
\label{fig:combined_pot}
\end{figure}



We show examples of the BO energy surface from independent phonon terms in
Fig.\ \ref{fig:combined_pot} for diamond and lithium hydride.  Both plots show
the BO energy surface as a function of phonon amplitude for an optical phonon
at the center of the Brillouin zone. The potential for diamond is asymmetric
and the dominant anharmonic term is cubic. Diamond has a face-centered cubic
(fcc) crystal structure with a two-atom basis at $(0,0,0)$ and
$(a/4,a/4,a/4)$, and a conventional cubic cell of lattice parameter $a$. The
optical phonon represents the motion of the two atoms of the basis in opposite
directions, leading to a steep potential when they approach, and a shallower
potential when they move apart. The lithium hydride crystal structure is also
fcc with a two-atom basis, but the second atom is located at $(a/2,a/2,a/2)$,
leading to a rather isotropic environment and a symmetric anharmonic
potential. In this case the anharmonicity arises because the potential is
steeper than the harmonic one, and the leading anharmonic term is quartic.


The crystal structure determines the symmetry operations relating different
$\mathbf{k}$-points within the first Brillouin zone. This implies that the
only points that need to be treated explicitly are those within the
irreducible wedge of the Brillouin zone, which reduces the computational cost
significantly. For example, a $5\times5\times5$ supercell of the diamond
structure has $125$ $\mathbf{k}$-points, of which only $22$ need be treated
explicitly.

\subsubsection{Vibrational self-consistent field}

Within the principal axes approximation to the BO energy surface the
vibrational Hamiltonian is
\begin{equation}
  \hat{H}_{\mathrm{vib}}=\sum_{n,\mathbf{k}}-\frac{1}{2}\frac{\partial^2}{\partial q^2_{n\mathbf{k}}} + E_{\mathrm{el}}(\mathbf{Q},\beta), \label{eq:anh_hamil}
\end{equation}
and the Schr\"{o}dinger equation is solved by an anharmonic vibrational wave
function $|\Phi(\mathbf{Q})\rangle$ and corresponding anharmonic energy
$E_{\mathrm{vib}}$, where
$\hat{H}_{\mathrm{vib}}|\Phi(\mathbf{Q})\rangle=E_{\mathrm{vib}}|\Phi(\mathbf{Q})\rangle$. In
Eq.\ (\ref{eq:anh_hamil}), $\beta$ is the inverse temperature. In principle it would be straightforward to calculate the BO energy surface with electrons distributed, for example, according to the Fermi-Dirac distribution at some finite temperature without substantial additional cost. This temperature dependence can be important for metallic systems, but we will drop it in the rest of this paper, because here we only study systems with a band gap.  

To solve this equation, the energy is minimized with respect to a set of
single-phonon states $\{|\phi_{n\mathbf{k}}(q_{n\mathbf{k}})\rangle\}$ that
form a Hartree product for the trial wave function
$|\Phi(\mathbf{Q})\rangle=\prod_{n,\mathbf{k}}|\phi_{n\mathbf{k}}(q_{n\mathbf{k}})\rangle$. This
leads to the vibrational self-consistent field equations\cite{bowman:608}
\begin{equation}
  \left(-\frac{1}{2}\frac{\partial^2}{\partial q_{n\mathbf{k}}^2}+\overline{V}_{n\mathbf{k}}(q_{n\mathbf{k}})\right)|\phi_{n\mathbf{k}}(q_{n\mathbf{k}})\rangle=\lambda_{n\mathbf{k}}|\phi_{n\mathbf{k}}(q_{n\mathbf{k}})\rangle, 
\end{equation}
where $\lambda_{n{\bf k}}$ is a vibrational energy eigenvalue and
\begin{equation}
\overline{V}_{n\mathbf{k}}(q_{n\mathbf{k}})\!=\!\left\langle\prod_{n',\mathbf{k}'}\!\!{}^{'}\phi_{n'\mathbf{k}'}(q_{n'\mathbf{k}'})\right|\!E_{\mathrm{el}}(\mathbf{Q})\left|\prod_{n',\mathbf{k}'}\!\!{}^{'}\phi_{n'\mathbf{k}'}(q_{n'\mathbf{k}'})\right\rangle.
\end{equation}
The primed products indicate that the term $(n,\mathbf{k})$ is excluded. The
final energy is
\begin{eqnarray}
&&E_{\mathrm{vib}}=\sum_{n,\mathbf{k}}\lambda_{n\mathbf{k}} + \\
&& \left\langle\prod_{n,\mathbf{k}}\phi_{n\mathbf{k}}(q_{n\mathbf{k}})\right|E_{\mathrm{el}}(\mathbf{Q})-\!\sum_{n,\mathbf{k}}\overline{V}_{n\mathbf{k}}(q_{n\mathbf{k}})\left|\prod_{n,\mathbf{k}}\phi_{n\mathbf{k}}(q_{n\mathbf{k}})\!\right\rangle. \nonumber
\end{eqnarray}

The VSCF equations determine a set of excited eigenstates for each degree of
freedom. These can be used to construct approximate anharmonic excited state
wave functions as
\begin{equation}
|\Phi^{\mathbf{S}}(\mathbf{Q})\rangle=\prod_{n,\mathbf{k}}|\phi^{S_{n\mathbf{k}}}_{n\mathbf{k}}(q_{n\mathbf{k}})\rangle,
\end{equation}
where the superindices $S_{n\mathbf{k}}$ label the excited states, and
$\mathbf{S}$ is a vector with elements $S_{n\mathbf{k}}$. The corresponding
energies are labeled by $E_{\mathbf{S}}$.

A perturbation theory can be constructed on the VSCF
equations\cite{norris:11261}, and the second order correction to the energy of
state $\mathbf{S}$ is
\begin{eqnarray}
&&E^{(2)}_{\mathrm{vib},\mathbf{S}}=\sum_{\mathbf{S}'\neq\, \mathbf{S}}\frac{1}{E_{\mathbf{S}}-E_{\mathbf{S}'}} \times \label {eq:mp2}\\
&&\left|\left\langle\prod_{n,\mathbf{k}}\phi_{n\mathbf{k}}^{S'_{n\mathbf{k}}}\right| E_{\mathrm{el}}(\mathbf{Q})-\sum_{n,\mathbf{k}}\overline{V}_{n\mathbf{k}}(q_{n\mathbf{k}})\left|\prod_{n,\mathbf{k}}\phi_{n\mathbf{k}}^{S_{n\mathbf{k}}}\right\rangle\right|^2\!\!. \nonumber
\end{eqnarray}

The anharmonic free energy $F$ can be calculated at any inverse temperature
$\beta$ from the eigenvalues that solve the VSCF equations as
\begin{equation}
F=-\frac{1}{\beta}\ln\mathcal{Z},
\end{equation}
where $\mathcal{Z}=\sum_{\mathbf{S}}e^{-\beta E_{\mathbf{S}}}$ is the partition function.

\section{Phonon expectation values} \label{sec:ph_theory}

\subsection{General formulation}

The wave function $|\Phi(\mathbf{Q})\rangle$ that solves the VSCF equations
contains, in principle, all of the physical information about the vibrational
system within the principal axes approximation. For an operator
$\hat{O}(\mathbf{Q})$ that depends on the phonon coordinates, it is then
possible to evaluate the expectation value with respect to the phonon wave
function as
\begin{equation}
\langle\hat{O}(\mathbf{Q})\rangle_{\Phi}=\langle\Phi(\mathbf{Q})|\hat{O}(\mathbf{Q})|\Phi(\mathbf{Q})\rangle.
\end{equation} 

The expansion of the operator $\hat{O}(\mathbf{Q})$ in terms of the phonon
amplitudes can be written in analogy to the expression for the BO
energy surface, as
\begin{eqnarray}
\hat{O}(\mathbf{Q})&=&\hat{O}(\mathbf{0})+\sum_{n,\mathbf{k}}\hat{O}_{n\mathbf{k}}(q_{n\mathbf{k}}) \nonumber \\
&+&\sum_{n,\mathbf{k}}\sum_{n',\mathbf{k}'}\hat{O}_{n\mathbf{k};n'\mathbf{k}'}(q_{n\mathbf{k}},q_{n'\mathbf{k'}})+\cdots
\end{eqnarray}
This expression can be constructed from data accumulated during the mapping of
the BO energy surface along the principal axes.

Within the finite temperature formalism, the expectation value of the
operator $\hat{O}(\mathbf{Q})$ at inverse temperature $\beta$ is
\begin{equation}
\langle\hat{O}(\mathbf{Q})\rangle_{\Phi,\beta}=\frac{1}{\mathcal{Z}}\sum_{\mathbf{S}}\langle\Phi^{\mathbf{S}}(\mathbf{Q})|\hat{O}(\mathbf{Q})|\Phi^{\mathbf{S}}(\mathbf{Q})\rangle
e^{-\beta E_{\mathbf{S}}}. \label{eq:ph_exp}
\end{equation}

Examples of phonon-dependent operators are electronic eigenvalues, the stress tensor, and the atomic positions. In this paper we will describe electronic eigenvalues and their relation to electron-phonon interactions and the stress tensor and its relation to thermal expansion.

\subsection{Electron-phonon interactions}

The effects of electron-phonon interactions on the electronic band structure
of solids can be calculated by renormalizing the phonons due to interactions
with the electronic states\cite{brooks,0295-5075-10-6-011} or renormalizing
the electrons due to their interactions with the
phonons.\cite{0022-3719-9-12-013} These two approaches can be shown to be
equivalent by Brooks' theorem.\cite{ZPBAllen}

We accumulate the single-particle electronic band structure while mapping the
BO energy surface within the principal axes approximation. Using the expansion
above for the phonon amplitude dependence, it is then possible to calculate
the renormalized electronic band structure due to the presence of the phonons.


\subsection{Stress tensor} \label{subsec:stress}

The differential Gibbs free energy $\mathrm{d}G$ of a system with an
externally applied stress $\sigma^{\mathrm{ext}}_{ij}$ is\cite{landau7}
\begin{equation}
\mathrm{d}G=\mathrm{d}F_{\mathrm{el}}+\mathrm{d}F_{\mathrm{vib}}-\Omega\sum_{i,j}\sigma^{\mathrm{ext}}_{ij}\mathrm{d}\epsilon_{ij}, \label{eq:vib_enthalpy}
\end{equation}
where $\epsilon_{ij}$ is the strain tensor and $\Omega$ is the volume. The
internal vibrational stress tensor is defined as 
\begin{equation}
\sigma^{\mathrm{vib}}_{ij}=-\frac{1}{\Omega}\frac{\partial F_{\mathrm{vib}}}{\partial\epsilon_{ij}}.
\end{equation}
The vibrational energy is an approximately linear function of volume for the
configurations of interest in thermal expansion. This leads to a constant
vibrational stress tensor. It is then possible to define an effective stress
$\sigma^{\mathrm{eff}}_{ij}=\sigma^{\mathrm{ext}}_{ij}+\sigma^{\mathrm{vib}}_{ij}$
such that Eq.\ (\ref{eq:vib_enthalpy}) can be rewritten as
\begin{equation}
\mathrm{d}G=\mathrm{d}F_{\mathrm{el}}-\Omega\sum_{i,j}\sigma^{\mathrm{eff}}_{ij}\mathrm{d}\epsilon_{ij}. \label{eq:elec_enthalpy}
\end{equation}
The minimum of the Gibbs free energy with respect to variations in strain
corresponds to the equilibrium configuration of the system. Therefore,
minimizing the Gibbs free energy in Eq.\ (\ref{eq:elec_enthalpy}), where the
vibrational dependence is implicit in the effective stress
$\sigma_{ij}^{\mathrm{eff}}$, leads to the equilibrium configuration for the
full vibrational system. If the vibrational stress varies significantly in the
region of interest, an iterative approach can be
adopted. Equation\ (\ref{eq:elec_enthalpy}) can be used to determine a new
equilibrium configuration, which can be used as a starting point for the next
calculation. This iterative process can be repeated until convergence is
reached. In this work, only one or two iterations were required for
convergence.

The stress tensor arising from the electronic Hamiltonian can be calculated
for each point sampled on the BO energy surface, and its expectation value
with respect to the phonon wave function evaluated using Eq.\ (\ref{eq:ph_exp}). This stress, $\sigma^{\mathrm{vib,V}}_{ij}$, is the contribution of the potential
energy to the vibrational stress. In general, the internal stress tensor has
contributions from both the kinetic and potential parts of the
Hamiltonian. The contribution from the kinetic part,\cite{PhysRevB.32.3780}
\begin{equation}
\sigma^{\mathrm{vib,T}}_{ij}=-\frac{1}{\Omega}\left\langle\Phi\left|\sum_{\mathbf{R}_p,\alpha}m_{\alpha}\dot{u}_{p\alpha;i}\dot{u}_{p\alpha;j}\right|\Phi\right\rangle,
\end{equation}
is therefore also required. The symbol $\dot{u}$ indicates the time derivative of a displacement coordinate. For an isotropic system where the internal vibrational stress can be described by a pressure $P$, the contribution from the kinetic part becomes $3P=-\mathrm{tr}\,\bm{\sigma}^{\mathrm{vib,T}}=2 E_{\mathrm{vib}}^{\mathrm{T}}/\Omega$, where $\mathrm{tr}$ denotes the trace of the tensor and $E_{\mathrm{vib}}^{\mathrm{T}}$ is the vibrational kinetic energy.

The total vibrational stress
$\sigma^{\mathrm{vib}}_{ij}=\sigma^{\mathrm{vib,V}}_{ij}+\sigma^{\mathrm{vib,T}}_{ij}$
can then be used in tandem with Eq.\ (\ref{eq:elec_enthalpy}) to determine the
equilibrium configuration in the presence of vibrations and at finite
temperature.


\section{Computational details} \label{sec:comp}

\subsection{Electronic calculations}

We have solved the electronic Schr\"{o}dinger equation within plane-wave
DFT\cite*{PhysRev.136.B864,PhysRev.140.A1133} using ultrasoft
pseudopotentials\cite{PhysRevB.41.7892} as implemented in the \textsc{castep}
code.\cite{CASTEP} All energy differences were converged to within $10^{-4}$
eV per unit cell and all stresses were converged to within $10^{-2}$ GPa.
This level of convergence required an energy cutoff of $1000$ eV and a
reciprocal-space Monkhorst-Pack\cite{PhysRevB.13.5188} grid of spacing $2\pi
\times 0.04$ \AA $^{-1}$.  We have used the local density approximation
(LDA)\cite*{PhysRevLett.45.566,PhysRevB.23.5048} to the exchange-correlation
functional for the diamond calculations, and the Perdew-Burke-Ernzerhof
(PBE)\cite{PhysRevLett.77.3865} generalized gradient approximation density
functional for the lithium hydride and deuteride calculations.

\subsection{Vibrational calculations}

We have mapped the BO energy surface along the phonon modes with a maximum
amplitude given by a multiple (between $3$ and $5$) of the harmonic
expectation value of the mode amplitude
\begin{equation}
\langle q^2_{n\mathbf{k}}\rangle=\frac{1}{\omega_{n\mathbf{k}}}\left(\frac{1}{2}+\frac{1}{e^{\beta\omega_{n\mathbf{k}}}-1}\right).
\end{equation}
We have fitted the independent phonon term with a polynomial of order $6$, and
tests with polynomials of order up to $8$ have confirmed the convergence of
the results. For the two-body terms we have fitted the two-parameter
functional form
$V_{n\mathbf{k};n'\mathbf{k}'}=c^{(1)}_{n\mathbf{k};n'\mathbf{k}'}\,q_{n\mathbf{k}}q_{n'\mathbf{k}'}
+
c^{(2)}_{n\mathbf{k};n'\mathbf{k}'}\,q^2_{n\mathbf{k}}q^2_{n'\mathbf{k}'}$. The
first term is the lowest order coupling term in a Taylor expansion of the
potential. The second term is not the next order term, but the lowest order term that gives a non-zero contribution when overlapped with the harmonic ground state, and, for this reason, it is expected to be larger than other lower order cross-terms. Numerical tests have confirmed this.

We use simple harmonic oscillator eigenstates as a basis set for expanding
the anharmonic wave function in the VSCF equations. The simple harmonic oscillator frequencies used are those of a quadratic fit to the anharmonic independent-phonon BO energy surface. We have used basis sets
including $100$ states for each mode, checking the convergence of the results
by using larger sets.

\section{Results} \label{sec:results}

\subsection{Anharmonic energy}

The mapping of the BO energy surface within the principal axes approximation
requires the displacement of the atomic nuclei inside a periodic
supercell. The supercell size determines a set of commensurate
reciprocal-space points associated with the harmonic phonon modes. Convergence
with respect to supercell size must be tested.

The sampling of the Brillouin zone within the HA is not constrained by the
supercell size in the same manner. This is because within the HA, all that is
needed is the dynamical matrix at the $\mathbf{k}$-points of interest, and
this matrix can be constructed at general $\mathbf{k}$-points irrespective of
the supercell size. This is accomplished by first constructing the real-space
matrix of force constants and then Fourier transforming to the dynamical
matrix at a general $\mathbf{k}$-point. This approach relies on the fact that
the force constants tend rapidly to zero as the separation between atoms
increases. The matrix of force constants can be constructed using a finite
displacement method\cite{PhysRevLett.40.950} or density functional
perturbation theory\cite*{PhysRevB.43.7231,RevModPhys.73.515} to evaluate the
dynamical matrices on a coarse $\mathbf{k}$-point grid, and then calculating the
matrix of force constants by means of an inverse Fourier transform.

The strategy we follow is to calculate both the harmonic $F_{\mathrm{har}}$
and anharmonic $F_{\mathrm{anh}}$ free energies from the $\mathbf{k}$-points
commensurate with the supercell. The anharmonic correction is then evaluated
as $\Delta F_{\mathrm{anh}}=F_{\mathrm{anh}}-F_{\mathrm{har}}$, and converged
with respect to supercell size. This correction is then added to an
independently converged harmonic free energy to give the total anharmonic
energy.

\begin{figure}
\includegraphics[scale=0.35]{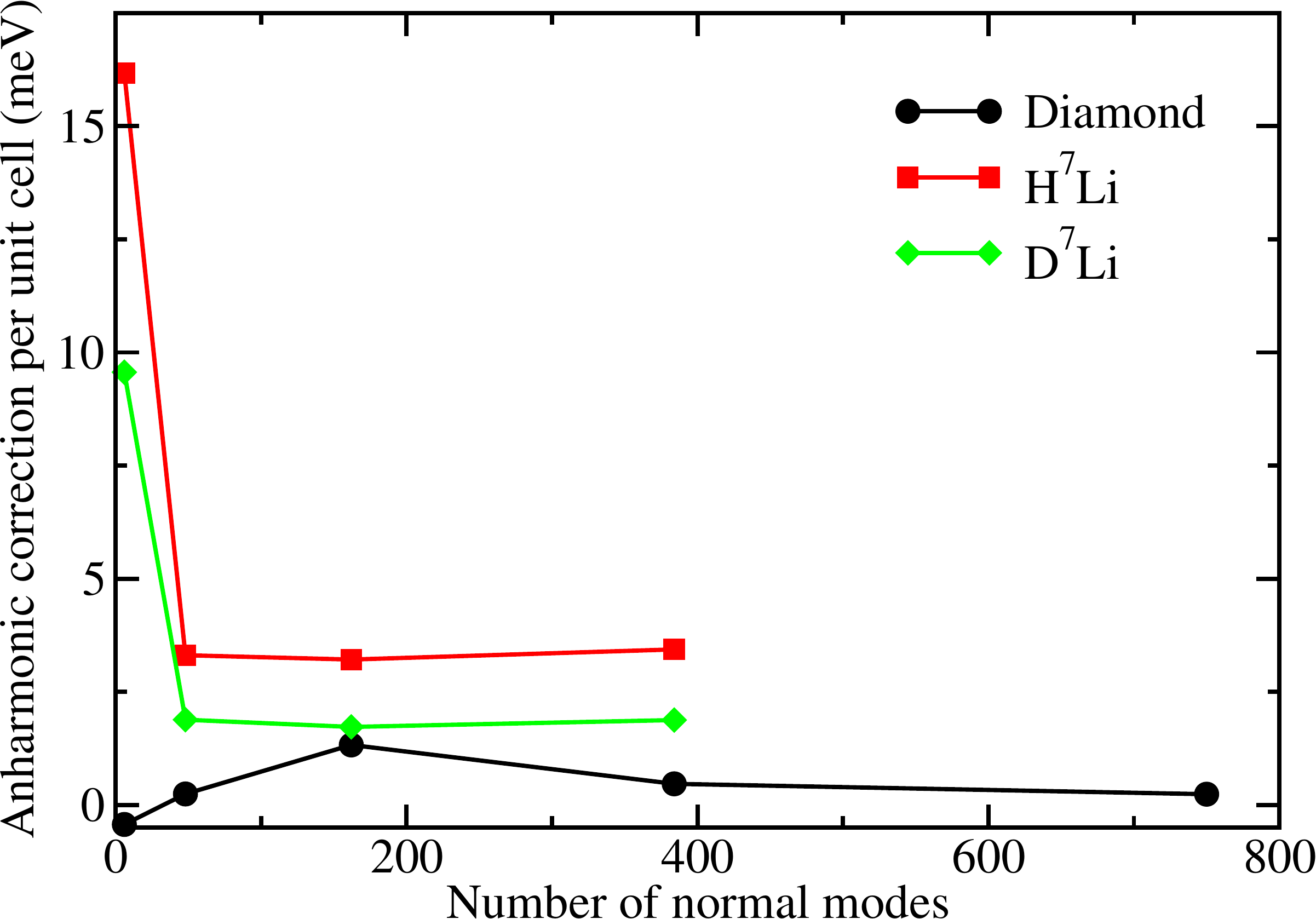}
\caption{(Color online) Anharmonic energy correction as a function of the
  number of normal modes (supercell size) for diamond, H$^7$Li and
  D$^7$Li. The anharmonic correction is larger for the lighter elements.}
\label{fig:sc_size}
\end{figure}

\begin{table}[b]
\caption{Converged anharmonic free energy correction per unit cell at zero
  temperature using independent phonons for the mapping of the BO energy
  surface along the principal axes. The converged harmonic energy is also
  displayed.}
\label{tab:anh_energy}
\begin{tabular}{lccc}
\hline
\hline
  & \hspace{0.1cm}  Num.\ modes \hspace{0.1cm} & $F_{\mathrm{har}}$ (meV) \hspace{0.1cm} &  $\Delta F_{\mathrm{anh}}$ (meV)  \\
\hline
Diamond & 750 & 367.59 & 0.23  \\
H$^7$Li & 384 & 221.49 & 3.21  \\
D$^7$Li & 384 & 174.75 & 1.88  \\
\hline
\hline
\end{tabular}
\end{table}

The supercell-size dependence of $\Delta F_{\mathrm{anh}}$ at zero temperature
for diamond, lithium hydride, and lithium deuteride is shown in
Fig.\ \ref{fig:sc_size}. The anharmonic energy is evaluated using only the
independent phonon terms in Eq.\ (\ref{eq:indep_term}). The energy difference
between the last two successive points in each curve is smaller than $0.2$ meV
per unit cell. Table \ref{tab:anh_energy} gives the numerical values of the
harmonic phonon energy and the independent phonon anharmonic correction per
unit cell.

The anharmonicity of diamond is small. For the smallest supercell size
corresponding to the $\Gamma$-point only, the anharmonic correction is
negative, as expected for cubic anharmonicity\cite{cohen_tannoudji2} (see
Fig.\ \ref{fig:combined_pot}). Including more $\mathbf{k}$-points reverses the
sign of the anharmonic correction, but it remains small.

The different isotopic masses of H$^7$Li and D$^7$Li lead to different vibrational properties. The anharmonic correction is larger for
H$^7$Li than for D$^7$Li because the lighter elements explore wider regions of
the BO energy surface and encounter more anharmonicity. We have also performed
calculations with the lithium isotope $^6$Li, but the results are not much
affected by this isotopic substitution because it represents a mass reduction
of only about $15$\%, compared to the doubling of mass from substituting
hydrogen by deuterium.

\begin{figure}
\includegraphics[scale=1.8]{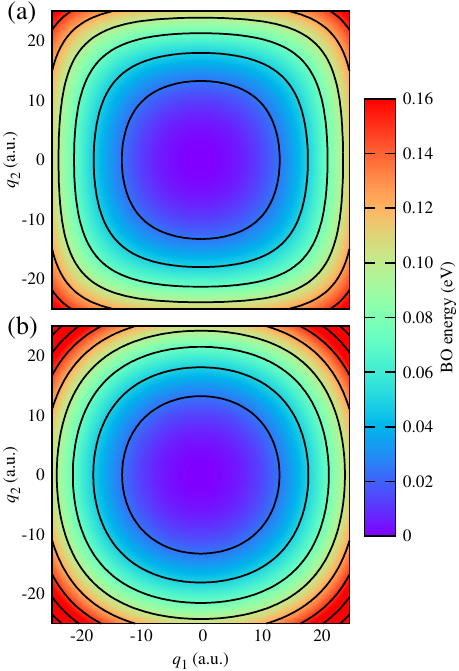}
\caption{(Color online) Anharmonic two-dimensional BO energy surface per unit
  cell for a pair ($q_1,q_2$) of optical modes at the $\Gamma$-point for
  H$^7$Li. We plot (a) the exact energy surface and (b) the approximate energy
  surface for independent phonons only. For a harmonic ground-state wave
  function (Gaussian), the one-sigma amplitude is $13.7$ a.u.}
\label{fig:2d_pot}
\end{figure}

We have studied the effects of including the phonon coupling term in
Eq.\ (\ref{eq:coup_term}) for a $2\times2\times2$ supercell ($48$ modes) of
H$^7$Li. The anharmonic coupling correction has the opposite effect to the
independent phonon correction, and the final anharmonic correction with
coupling between modes is $\Delta F_{\mathrm{anh}}=+0.79$ meV, compared to
$\Delta F_{\mathrm{anh}}=+3.21$ meV with independent phonons only (see
Table\ \ref{tab:anh_energy}). Using the second order perturbation theory of
Eq.\ (\ref{eq:mp2}) does not change the final result appreciably, indicating that
the energies have converged with respect to the mean-field theory. This means
that, like diamond, H$^7$Li has a small anharmonic energy correction, but, unlike diamond,
the reason for this is the cancellation of the contributions from independent
phonons and pairwise phonon coupling. This negative contribution from the
coupling terms can be better appreciated by referring to
Fig.\ \ref{fig:2d_pot}. We show the coupling BO energy surface
$V_{1;2}(q_1,q_2)$ for two optical phonons ($q_1,q_2$) at the $\Gamma$-point
compared to the approximate surface constructed from the independent phonon
terms only. The accurate potential is shallower than the approximate one,
leading to the smaller anharmonic correction when including phonon
coupling. The corresponding harmonic two-dimensional subspace is
very similar to the approximate (bottom) potential in Fig.\ \ref{fig:2d_pot},
but with concentric circular equipotential lines because the harmonic phonon
frequencies are degenerate in the case shown.

Nolan and co-workers\cite{PhysRevB.80.165109} have used very accurate
electronic structure methods to calculate the cohesive energy of lithium
hydride, including the zero point (ZP) energy at the harmonic level. Their
accuracy is of the same order of magnitude as the anharmonic corrections we
have calculated for this system.

Figure \ref{fig:sc_size} shows that convergence with respect to supercell size
is reached with smaller cells for H$^7$Li and D$^7$Li than for diamond. In
diamond, the phonons with the largest anharmonic contribution to the energy
are located in the third phonon branch, corresponding to the highest energy
acoustic modes. These modes represent large-scale atomic displacements and
larger supercells are needed to describe them. In contrast, the largest
anharmonic contributions in H$^7$Li and D$^7$Li arise from the optical
branches, which represent small scale displacements of the atoms within the
primitive cells. This local anharmonic behavior is reflected by convergence at
smaller supercell sizes.

\subsection{Band gap renormalization}

We have used the phonon expectation value formalism to investigate the ZP
renormalization and temperature dependence of band gaps in diamond and
isotopes of lithium hydride. In both cases we have selected the thermal band
gap (smallest band gap), which is indirect for diamond and direct for the lithium
hydride systems. For diamond, the top of the valence band is at the electronic
$\Gamma$ point, and the bottom of the conduction band is along the line
joining the $\Gamma$ and X points. The thermal gap is at the electronic X
point for the lithium hydride isotopes. In all calculations reported in this
section the anharmonicity has been treated at the independent phonon level.

\begin{figure}
\includegraphics[scale=0.35]{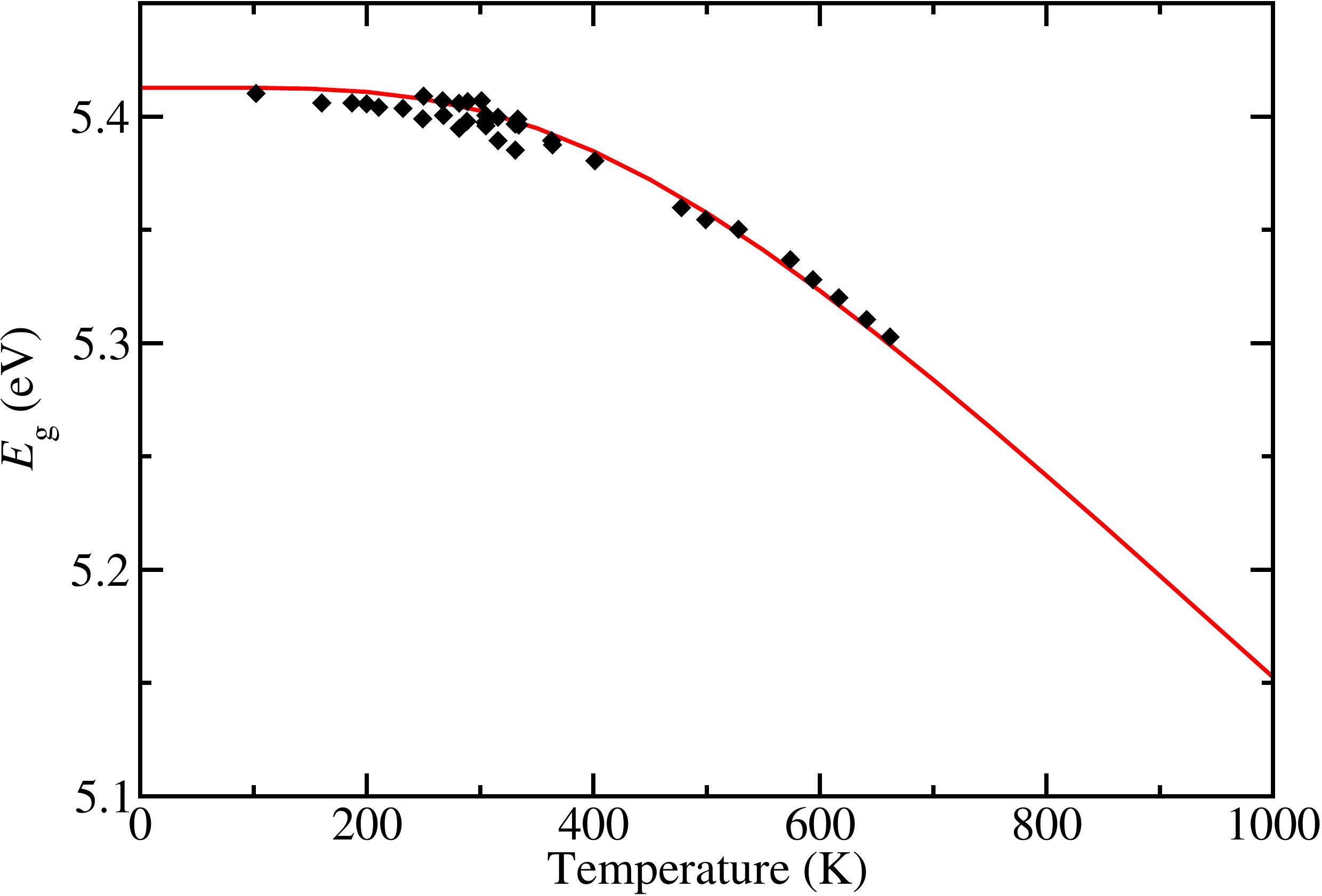}
\caption{(Color online) Temperature dependence of the thermal band gap $E_{\mathrm{g}}$
  of diamond. The DFT result (red solid curve) is offset to match experimental data (black diamonds) at zero temperature. The experimental data are from Ref.\ \onlinecite{Clark11021964}. } 
\label{fig:diamond_bg}
\end{figure}

We show the temperature dependence of the electronic thermal band gap
$E_{\mathrm{g}}$ of diamond in Fig.\ \ref{fig:diamond_bg}. The black diamonds
are experimental results from Ref.\ \onlinecite{Clark11021964}. The ZP band
gap renormalization is $E^{\mathrm{ZP}}_{\mathrm{g}}=-462$ meV\@. This value
is $92$ meV larger than that estimated from the asymptotic behavior at high
temperature in Ref.\ \onlinecite{Cardona20053}. The analysis in
Ref.\ \onlinecite{Cardona20053} suffers from the lack of experimental data
beyond the Debye temperature, where the asymptotic behavior should occur, and
an analytic functional form with the correct asymptotic behavior was fitted to the data instead. We attribute some of
the discrepancy to the inaccuracy of the fitted function used in
Ref.\ \onlinecite{Cardona20053}. Our first-principles results can also be compared with other theoretical studies of the thermal gap of diamond\cite{PhysRevB.45.3376} based on the empirical pseudopotential method\cite{PhysRev.141.789} which find a ZP renormalization of approximately $-620$ meV. The calculations reported in Ref.\ \onlinecite{PhysRevB.45.3376} lead to an overestimate of the reduction in band gap with temperature compared to experiment. This could explain the larger ZP renormalization found in Ref.\ \onlinecite{PhysRevB.45.3376} compared to our first-principles results. The band gap renormalization has an
additional intrinsic contribution from the change in volume of the crystal
with temperature. For diamond this contribution is very small, amounting to a
decrease of $16$ meV for the temperature range in Fig.\ \ref{fig:diamond_bg}.

\begin{figure}
\includegraphics[scale=0.35]{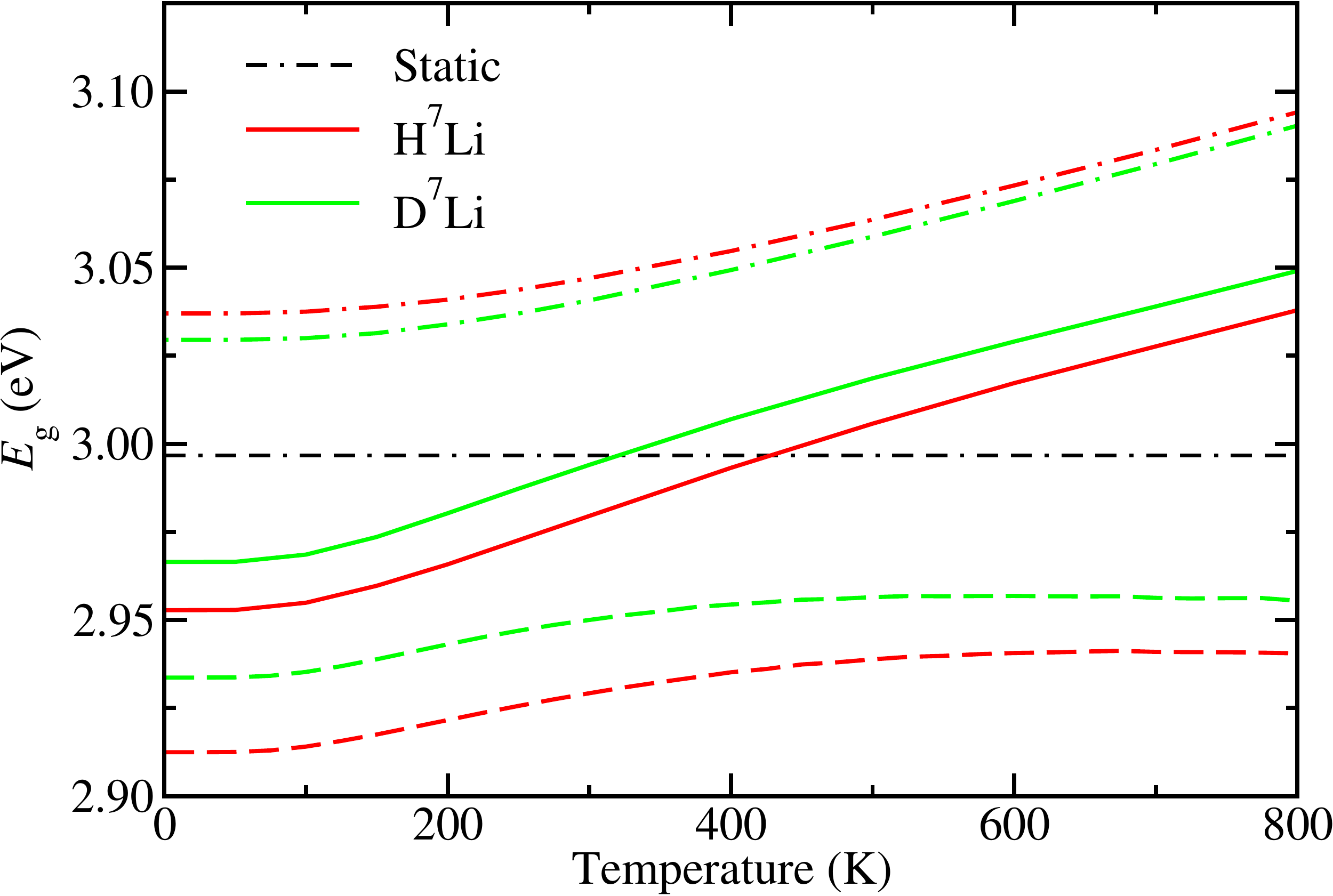}
\caption{(Color online) Temperature dependence of the thermal band gap $E_{\mathrm{g}}$
  of lithium hydride for the isotopes H$^7$Li (red) and D$^7$Li (green) calculated within DFT\@. The black double-dashed-dotted line indicates the value of the static band
  gap. Dashed lines refer to the renormalized gaps due to electron-phonon
  interactions, dashed-dotted lines refer to the renormalization due to
  thermal expansion with temperature, and the solid lines are the sum of the
  two contributions.}
\label{fig:lih_bg}
\end{figure}

Figure\ \ref{fig:lih_bg} shows the temperature dependence of the thermal band
gaps of H$^7$Li and D$^7$Li. The ZP renormalizations arising from
electron-phonon interactions are $E^{\mathrm{ZP}}_{\mathrm{g}}=-84$ meV and
$E^{\mathrm{ZP}}_{\mathrm{g}}=-62$ meV, respectively; and those arising from
the change in volume with temperature are $E^{\mathrm{ZP}}_{\mathrm{g}}=+40$
meV and $E^{\mathrm{ZP}}_{\mathrm{g}}=+33$ meV, respectively. The isotope
effect for the lighter compound leads to a larger band gap renormalization for
both electron-phonon and volume contributions. This is in agreement with the
expectation that lighter isotopes show larger atomic displacements. 

It is interesting to note in the lithium hydride systems that the ZP electron-phonon renormalizations are
negative (gap closing), while opening of the gap is observed at finite
temperature. To investigate this we have calculated the contribution to the
electron-phonon renormalization from each phonon mode, and the results
indicate that low-energy phonons tend to open the gap whereas high-energy
phonons tend to close it. At the ZP level, all phonons contribute, and the
high energy gap-closing phonons dominate. However, as the temperature is
increased, the lower energy phonons are excited more easily, leading to
opening of the gap. At higher temperatures, when the high-energy phonons are
excited as well, the trend is expected to revert to gap closing. In
Fig.\ \ref{fig:lih_bg}, the higher temperatures explored (close to the melting
point) only show a leveling off of the electron-phonon correction. Extending
the calculations to higher unphysical temperatures (not shown) does take the
system into the gap-closing regime as expected.

Our first-principles calculations for lithium hydride are, as far as we are
aware, the first to show nonmonotonic behavior in the temperature dependence
of a band gap due to electron-phonon interactions. This behavior has been
observed experimentally in ternary
semiconductors\cite{Artus1987733,PhysRevB.86.195208} and it has been described
by an analytic functional form consisting of the sum of two Bose-Einstein
oscillators of opposite sign and with energies related to the low and high
frequency phonon modes.\cite{PhysRevB.86.195208} Our calculations show that
this difference in behavior between the low and high frequency phonons is
responsible for the nonmonotonic variation of the gap with temperature.

In Fig.\ \ref{fig:lih_bg} we also show the sum of the contributions to the gap
renormalization from electron-phonon interactions and thermal expansion for H$^7$Li and D$^7$Li. This
simple addition of the two terms is the standard procedure used in the
literature.\cite{PhysRevB.23.1495} However, this approximation might not be
very accurate for a system such as lithium hydride with a large thermal
expansion (see Sec.\ \ref{subsec:thermal_exp} below). We calculated the
electron-phonon interaction renormalization to the band gap at the volume
including ZP motion for H$^7$Li, and the real renormalization is $4$ meV
larger than the one found by the simple addition procedure.

In Fig.\ \ref{fig:diamond_bg}, a constant energy shift has been added to the
theoretical curve to match experiment at zero temperature. Standard
approximations to the DFT exchange-correlation functional such as the LDA or
PBE do not reproduce the derivative discontinuity with respect to particle
number of the exact functional. This leads to a severe underestimation of band
gaps in semiconductors and
insulators.\cite*{PhysRevLett.56.2415,PhysRevB.37.10159} This problem can be
addressed in various ways, the most straightforward being the application of a
scissor operator\cite*{PhysRevLett.56.2415,PhysRevLett.62.2160} to the band
gaps.
When calculating band-gap renormalization, we are interested in the
\textit{change} in the band gap, rather than its absolute value. The fact that
all atomic configurations suffer from a similar gap underestimation suggests
that the calculated changes may be accurate even when using standard
functionals. This is supported by the calculations of Giustino and
co-workers\cite{PhysRevLett.105.265501} of the electron-phonon band-gap
renormalization in diamond using Allen-Heine-Cardona perturbation
theory.\cite*{0022-3719-9-12-013,PhysRevB.23.1495} They use both the LDA and
the LDA corrected with a scissor operator, obtaining consistent results. Note
that Giustino and co-workers look at the direct optical gap instead of the
thermal gap that we have studied. Cannuccia and
Marini\cite{PhysRevLett.107.255501,cannuccia_ejpb} use many-body perturbation
theory to study the optical gap of diamond as well, and they report a
significant structure in the spectral function beyond the quasiparticle
picture, corresponding to subgap polaronic states.  

\begin{figure}
\includegraphics[scale=0.35]{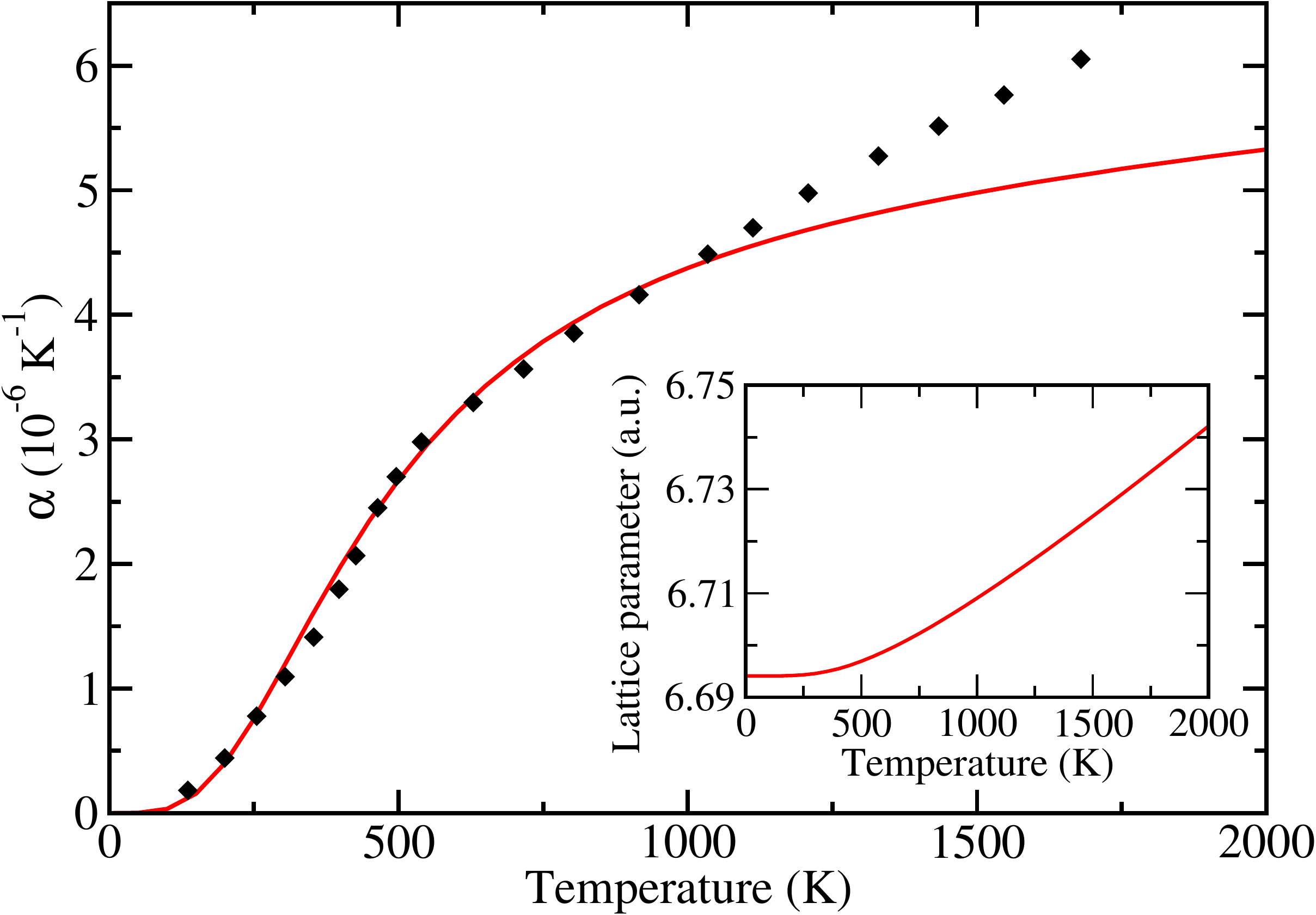}
\caption{(Color online) Temperature dependence of the coefficient of linear
  expansion $\alpha$ for diamond. The black diamonds are experimental results
  from Ref.\ \onlinecite{slack:89}. The inset shows the temperature dependence
  of the lattice parameter.}
\label{fig:diamond_th}
\end{figure}

\subsection{Thermal expansion} \label{subsec:thermal_exp}

We have used the formalism described in Sec.\ \ref{subsec:stress} to calculate
the internal vibrational stress and the corresponding equilibrium
configuration of the crystal as a function of temperature. The structures of
diamond, H$^7$Li, and D$^7$Li are described by a single parameter, so the
different equilibrium configurations differ only in the volume. We present
results for the thermal expansion of diamond, H$^7$Li, and D$^7$Li. All
calculations reported in this section have treated anharmonicity at the
independent-phonon level.  

In Fig.\ \ref{fig:diamond_th} we show the temperature dependence of
the lattice parameter $a(T)$ of diamond and the corresponding
temperature dependence of the linear coefficient of thermal expansion,
\begin{equation}
\alpha(T)=\frac{1}{a}\frac{\mathrm{d}a}{\mathrm{d}T}.
\end{equation}
We compare our results with experimental data from
Ref.\ \onlinecite{slack:89}. The agreement is good for temperatures below
$1200$ K, but at higher temperatures the calculated $\alpha(T)$ underestimates
the experimental values. This effect is also seen in calculations using the
quasiharmonic approximation.\cite{PhysRevB.71.205214} Our methodology assumes
a volume-independent internal vibrational stress tensor for the volumes of
interest at each iteration, as described in Sec.\ \ref{subsec:stress}. 
Iterating the calculation at the volume including the ZP phonon pressure,
leads to a change in the phonon pressure of diamond of less than $-0.05$~GPa,
which represents less than $1.5$\% of the total phonon pressure. This means
that for diamond a single iteration is sufficient for obtaining converged
results.

\begin{table}[b]
  \caption{Static DFT and ZP renormalized lattice parameters of diamond,
    H$^7$Li, and D$^7$Li.}
\label{tab:latt_param}
\begin{tabular}{lccc}
\hline
\hline
  & \hspace{0.2cm}  $a^{\mathrm{static}}$ (a.u.) \hspace{0.2cm} &  $a^{\mathrm{ZP}}$ (a.u.) & $\Delta a$ (a.u.) \\
\hline
Diamond (LDA) & 6.669  & 6.694  &  +0.025 \\
H$^7$Li (PBE) & 7.600  & 7.758  &  +0.158 \\
D$^7$Li (PBE) & 7.600  & 7.726  &  +0.126 \\
\hline
\hline
\end{tabular}
\end{table}

\begin{figure}
\includegraphics[scale=0.35]{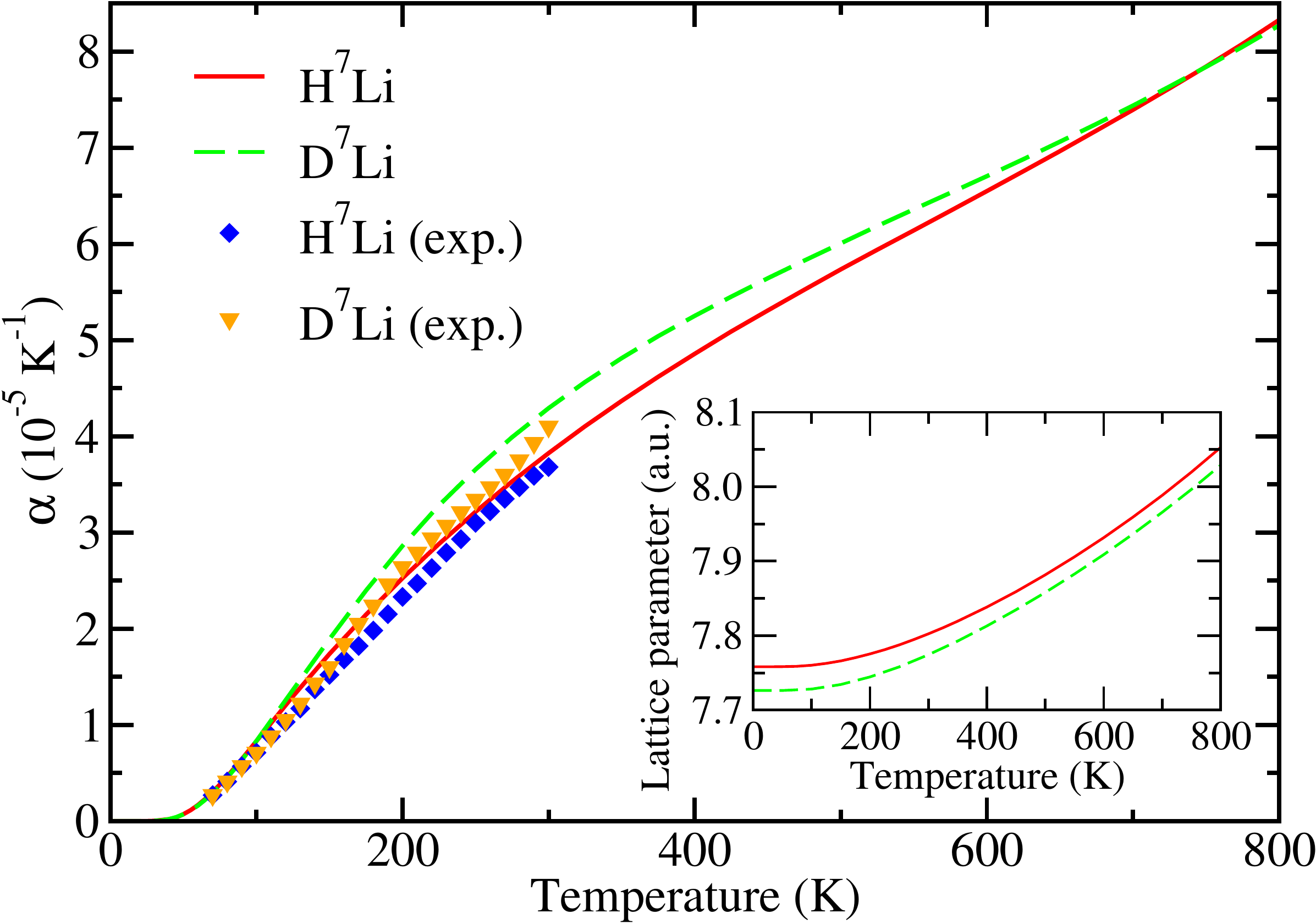}
\caption{(Color online) Temperature dependence of the coefficient of linear
  expansion $\alpha$ for lithium hydride with isotopic compositions H$^7$Li
  and D$^7$Li. The experimental data are from
  Ref.\ \onlinecite{0022-3719-15-31-009}. The inset shows the temperature
  dependence of the lattice parameters.}
\label{fig:lih_th}
\end{figure}

The results for the lithium hydride isotopes are shown in
Fig.\ \ref{fig:lih_th} and are compared with experimental data from
Ref.\ \onlinecite{0022-3719-15-31-009}. The lattice parameters of the
different isotopes show the expected behavior, with the lighter compound
leading to a larger ZP phonon pressure and therefore to a larger lattice
expansion. In contrast, the linear coefficient of thermal expansion is larger
for the heavier deuterium compound, in agreement with experiment. The phonons
of D$^7$Li have lower energies than those of H$^7$Li, and therefore they are
excited more easily with increasing temperature. This leads to the thermal
expansion coefficient being larger for D$^7$Li at low temperatures.

The volume change in lithium hydride due to ZP motion is much larger than that
in diamond. The change in ZP phonon pressure between the first and second
iterations is as large as $-0.22$ GPa, representing $11$\% of the absolute
value. In this case the second iteration is important for obtaining accurate
results.  The change in pressure is negative because the expanded lattice
(after the first iteration) has softer phonons than the static DFT lattice,
which leads to smaller phonon pressures.

The absolute values of the lattice parameters within DFT are known to deviate
systematically by a few percent from experimental results. However, the
differences in lattice parameters within a given DFT approximation are
expected to be accurate. These are displayed in Table
\ref{tab:latt_param}. Mounet and Marzari
(Ref.\ \onlinecite{PhysRevB.71.205214}) find the same ZP expansion for diamond
using the PBE functional, rather than the LDA that we have used, even though
the absolute value of the lattice parameter is somewhat different. Anderson
and co-workers\cite{0022-3719-15-31-009} report experimental results for the
lattice parameter of H$^7$Li and D$^7$Li at a temperature of $83$ K, finding a
difference between the value for the two isotopes of $0.032$ a.u., in good
agreement with the value of $0.034$ a.u.\ we have calculated.




\section{Conclusions} \label{sec:conclusions}

We have described a formalism for studying temperature dependent anharmonic
vibrational properties in periodic systems using first-principles
calculations. The BO energy surface is mapped using a principal axes
approximation, and the resulting equations are solved using a Hartree
mean-field approach and second-order perturbation theory. We then use the
vibrational anharmonic wave function to calculate expectation values of
phonon-dependent quantities. In particular, we have studied the effects of
electron-phonon interactions on electronic band gaps, and the role of the
stress tensor in thermal expansion. Other quantities can be studied within
this framework, such as the mean atomic positions within the unit cell.

The methodology we have presented involves the solution of the electronic
Schr\"{o}dinger equation at a set of low energy atomic configurations. Our
results are based on density functional theory calculations, but the
methodology is independent of the particular electronic structure method
used. For instance, for larger systems it might be necessary to use force
fields, or, if higher accuracy is required, quantum Monte Carlo
methods.\cite{RevModPhys.73.33}

We have tested the formalism on diamond, lithium hydride and lithium
deuteride, which exhibit quite small anharmonicities.  However, while diamond
is nearly harmonic, the small anharmonicity in lithium hydride arises from a
cancellation of the contributions from single-phonon and two-phonon terms. The
results for both band-gap renormalization and lattice expansion as a function
of temperature have been found to be in good agreement with experimental
results for diamond. For H$^7$Li and D$^7$Li, our thermal expansion results
also agree with experiment but, as far as we are aware, there are no
experimental data for the temperature dependence of the band gap and our results serve as a prediction that could be tested.

In our calculations using the VSCF method plus perturbation theory we appear
to have converged the results with respect to the description of
anharmonicity.  This suggests that our approach can be used to estimate
anharmonic effects in systems with substantially stronger anharmonicity than
diamond and lithium hydride.


\begin{acknowledgments}
  We thank the Engineering and Physical Sciences Research Council (UK)
  for financial support.  The calculations were performed on the
  Cambridge High Performance Computing Service facility. 
\end{acknowledgments}

\bibliography{anharmonic}

%
%
%
%
%
%

\end{document}